\documentclass[aps,pra,reprint,twocolumn,nofootinbib,superscriptaddress,showpacs,showkeys,superscriptaddress,longbibliography]{revtex4-1}
\usepackage{amsmath,amssymb,amstext}
\usepackage[usenames,dvipsnames]{color}
\usepackage{graphicx}
\usepackage{bm,bbold,braket}
\usepackage{natbib}
\usepackage{txfonts, comment,stmaryrd}
\usepackage{dcolumn}
\usepackage{color}
\usepackage{xcolor}
\colorlet{rn}{red}

\colorlet{gc}{blue}

\usepackage[english]{babel}
\usepackage[colorlinks,bookmarks=false,citecolor=blue,linkcolor=red,urlcolor=blue]{hyperref}
\begin{document}

\title{Commensurate supersolids and re-entrant transitions in an extended Bose-Hubbard ladder}
\author{Ashwath N Madhusudan}
\thanks{These authors contributed equally to this work.}
\affiliation{Department of Physics, Indian Institute of Science Education and Research, Pune 411 008, India}  
\affiliation{
Universit\"{a}t Innsbruck, Fakult\"{a}t f\"{u}r Mathematik, Informatik und Physik, Institut f\"{u}r Experimentalphysik, 6020 Innsbruck, Austria
} 
\author{Gopal Chandra Santra}
\thanks{These authors contributed equally to this work.}
 \affiliation{Department of Physics, Indian Institute of Science Education and Research, Pune 411 008, India} 
  \affiliation{Kirchhoff-Institut f\"ur Physik, Universit\"at Heidelberg, Im Neuenheimer Feld 227, 69120 Heidelberg, Germany}
      \affiliation{Pitaevskii BEC Center and Department of Physics, University  of  Trento,  Via Sommarive 14, I-38123 Trento, Italy}
       \affiliation{INFN-TIFPA, Trento Institute for Fundamental Physics and Applications, Via Sommarive 14, I-38123 Trento, Italy}
 \author{Inderpreet Kaur}
\affiliation{Department of Physics, Indian Institute of Science Education and Research, Pune 411 008, India}
\author{Weibin Li}
\affiliation{School of Physics and Astronomy, University of Nottingham, Nottingham, NG7 2RD, United Kingdom}
\author{Rejish Nath}
\affiliation{Department of Physics, Indian Institute of Science Education and Research, Pune 411 008, India}

\begin{abstract}
We investigate the ground state phases of an extended Bose-Hubbard ladder of unit filling via the density-matrix-renormalization-group method and, in particular, the effect of rung-hoppings. In contrast to a single-chain, a commensurate supersolid emerges, and based on the Luttinger parameter, we classify them into two types. The latter leads to a reentrant gapless behavior as the onsite interaction is increased while keeping all other parameters intact. A reentrant gapped transition is also found as a function of nearest-neighbor interactions. Further, we show that the string order characterizing the Haldane phase vanishes for a finite inter-chain hopping amplitude, however small it is. Finally, we propose two experimental platforms to observe our findings, using either dipolar atoms or polar molecules and Rydberg admixed atoms.
\end{abstract}

\maketitle


\section{Introduction}

The Bose-Hubbard model is characterized by onsite two-body interactions with hopping and undergoes a phase transition from the gapless superfluid (SF) to a gapped Mott insulator (MI) as the interaction strength is increased compared to the hopping amplitude \cite{fis89,lew07,blo12, lew12,kru16}. In the extended Bose-Hubbard model (EBHM) in which the nearest-neighbor interactions also present, interesting phases like supersolid (SS), density wave (DW), phase-separation (PS) and Haldane insulator (HI) emerge \cite{kuh00,pai05, lan16, bar12,sou19, stu20, bat06,niy94,bat15,mis09}. MI and SF have uniform densities, whereas SS and DW exhibit modulated occupation patterns. The exotic gapped HI is characterized by a nonlocal string order parameter \cite{dal06, ber08}. The lattice geometry and filling fraction $\rho$, the ratio between the number of bosons and lattice sites, also play a decisive role in the physics of ground-state phases. 

For weak onsite interactions, the ground state phase diagram of one-dimensional (1D) EBHM with unit filling ($\rho=1$) comprises three phases: SF, PS, and DW, respectively, at low, intermediate, and large values of nearest neighbor interactions. As onsite interactions increase, HI emerges, and PS is no longer the ground state. For large onsite interactions, the gapless SF phase is replaced by the gapped MI \cite{dal06, ber08, jam14, eji14, den11, stu20, den13, ros12}. Pure supersolids, possessing simultaneous superfluid and crystalline order, are not conclusively reported in $\rho=1$ ground state phase diagrams. The flavor of supersolidity for unit filling ($\rho=1$) emerges at small onsite interactions via the PS phase where the uniform SF is in the bulk and DW character at the edges \cite{bat14, kot21}. The commensurate 1D SSs are reported for high filling fractions such as $\rho=2$ and  $\rho=3$ \cite{bat13, bat14} and are expected to exist for integer fillings with $\rho>3$. In higher dimensions, the commensurate SSs exist for low fillings such as $\rho=1$ and $\rho=1/2$ \cite{mel05,hei05,wes05,sou19,ohg12, ohg12b, isk11, kim11,bog19}. The incommensurate SSs are seen by doping a DW \cite{bat06, niy94, bat15, mis09, kuh98}.

\begin{figure}
    \centering
    \includegraphics[width=\columnwidth]{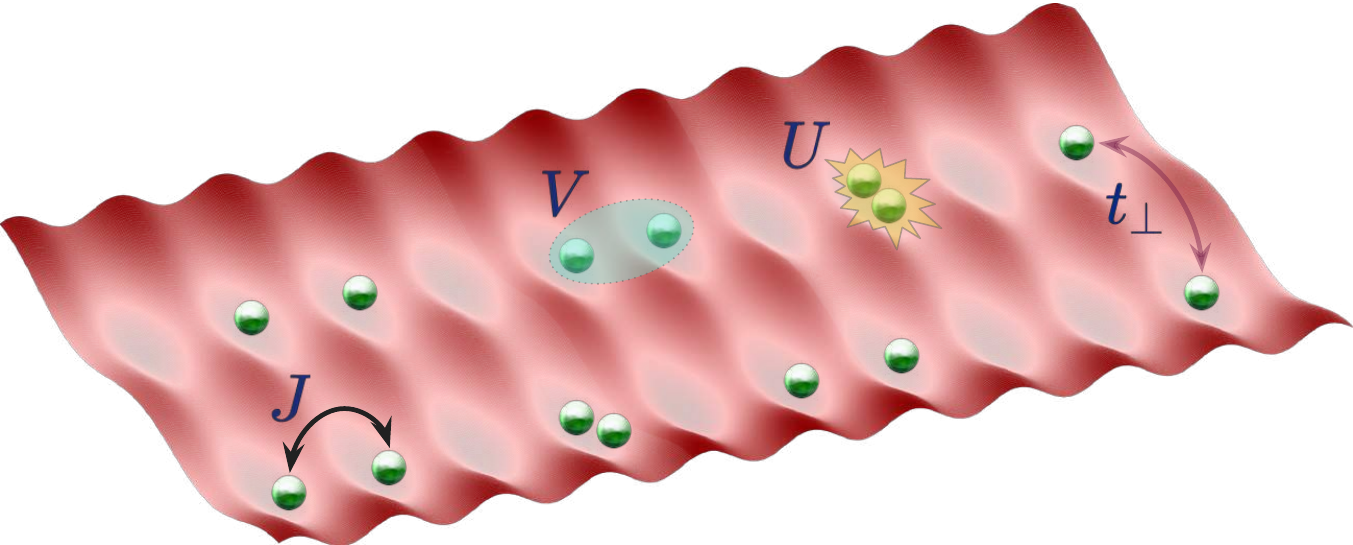}
    \caption{(Color online) The schematic diagram for the ladder setup. $J$ is the strength of the hopping along each chain, $V$ is the nearest-neighbor interaction in each chain, $U$ is the onsite interaction strength, and $t_{\perp}$ is the amplitude of the inter-chain or rung hopping.} 
    \label{fig:1}
\end{figure}

In this paper, we study the effect of rung-hoppings on the ground state phases of a two-leg bosonic ladder \cite{ber08,lee08,cre11,sin14,don23} of unit-filling using the finite-size density-matrix-renormalization-group (DMRG) method. The quasi-one-dimensional ladder can exhibit qualitatively different physics compared to the 1D counterpart. The density-density interactions between the two chains are disregarded in our study, which is a source of other interesting phases such as pair superfluidity, etc. \cite{ori98,arg07,pan15, sin17}. In the absence of nearest-neighbor interactions, a unit-filled Bose-Hubbard ladder exhibits a transition from MI to SF when the rung-hoppings are increased \cite{lut08, don01}, indicating an enhancement of delocalization of bosons within a chain. Counter-intuitively, the rung-hoppings can also introduce a Mott gap at half-fillings (rung-MI), which is found for both hardcore \cite{cre11,car11} and softcore bosons \cite{den15, ash23}. These findings indicate that the rung hoppings generally make the interplay between the hopping and inter-particle interactions within a chain non-trivial. As we include the nearest neighbor interactions, ground state features change drastically. In particular, we show that even relatively small rung hoppings can lead to a commensurate SS ground state in a unit-filled Bose-Hubbard ladder, in contrast to the one-dimensional case. It paves the way for a gapped-gapless transition, i.e., from DW to SS, as the rung hopping amplitude increases, similar to the MI-SF transition in the usual Bose-Hubbard ladder. The rung hoppings can also induce density wave character, resulting in a transition from SF to SS via PS at low onsite interactions. PS is found to be robust against inter-chain hoppings. Generally, the density wave order parameter shows a non-trivial dependence on the rungs hopping amplitude. Based on the value of the Luttinger parameter, we classify the supersolid into two types. Strikingly, our studies reveal reentrance behaviors, particularly reentrant gapless and gapped transitions. We also show that the string order vanishes for arbitrarily small rung hopping amplitudes, i.e., the Haldane phase is unstable against an arbitrarily small inter-chain hopping. Finally, we discuss two experimental platforms where our findings can be probed.

The paper is structured as follows. In Sec.~\ref{sm}, we introduce the model and the correlation functions characterizing different quantum phases. The ground state phases of the EBHM ladder are discussed in Sec.~\ref{gsp}. In particular, for small contact interaction in Sec.~\ref{u08}, intermediate in Sec.~\ref{u2} and sufficiently large in Sec.~\ref{loi}. The re-entrant transitions are discussed in Sec.~\ref{rent}. Finally, in Sec.~\ref{exp}, we discuss two experimental platforms where the model can be realized. The summary and outlook are provided in Sec.~\ref{cao}.


\section{Hamiltonian and correlation functions}
\label{sm}
We consider a two-leg Bose-Hubbard ladder in which each chain has $N$ bosons and $L$ sites and is described by the Hamiltonian, 
\begin{align}
H_J=-J\sum_{i=1}^{L}\sum_{\alpha=1, 2}( \hat b^\dagger_{i,\alpha}\hat b_{i+1,\alpha}+{\rm H.c.})-t_\perp\sum_{i=1}^{L}( \hat b^\dagger_{i,1}\hat b_{i,2}+ \rm{H.c.})\nonumber \\
+\frac{U}{2}\sum_{i=1}^{L}\sum_{\alpha=1, 2}\hat n_{i,\alpha}(\hat n_{i,\alpha}-1)+V\sum_{i=1}^{L-1}\sum_{\alpha=1, 2}\hat n_{i,\alpha }\hat n_{i+1,\alpha},
\label{ham}
\end{align}
where $\hat b^\dagger_{i,\alpha} ( \hat b_{i,\alpha})$ is the bosonic creation (annihilation) operator and $\hat n_{i, \alpha}= \hat b^\dagger_{i,\alpha} \hat b_{i,\alpha}$ is the number operator at the site $i$ of $\alpha$th chain. The first two terms in Eq.~(\ref{ham}) account for the particle hopping along the chain and in the rungs, respectively, with amplitudes $J$ and $t_\perp$. The last two terms are onsite and nearest-neighbor two-particle interaction Hamiltonians having strengths $U$ and $V$, respectively. We fix the filling fraction to $\rho=N/L=1$ unless otherwise specified and use open boundary conditions. We use the finite-size DMRG method to study the ground state phases, and the phase boundaries are obtained by finite-size scaling. The DMRG calculations are performed using the TeNPy Library \cite{tenpy}. We use a bond dimension of 800, except for Fig.~\ref{fig:2} and a maximum number of bosons per site of $n_b=6$, ensuring the results converged.


To identify different ground state phases, we use the following correlation functions calculated in each chain, 
\begin{eqnarray}
C_{{\rm SF}}(r)=\langle \hat b_{j, \alpha}^{\dagger}\hat b_{j+r, \alpha}\rangle, \\ 
C_{{\rm DW}}(r)=(-1)^r\langle \delta\hat n_{j, \alpha}\delta\hat n_{j+r, \alpha}\rangle, \\
C_{{\rm St}}(r)=\langle \delta\hat n_{j, \alpha}e^{i\pi\sum_{k=j}^{k=j+r-1}\delta\hat n_{k, \alpha}}\delta\hat n_{j+r, \alpha}\rangle,
\end{eqnarray}
where $\delta \hat n_{j, \alpha}=\hat n_{j, \alpha}-\rho$. $C_{{\rm SF}}(r)$, $C_{{\rm DW}}(r)$, and $C_{{\rm St}}(r)$ are SF, DW, and string correlation functions, respectively. The Haldane insulator is identified by a finite string order parameter, $\mathcal O_{{\rm St}}=\lim_{r\to\infty}C_{{\rm St}}(r)$. The SF phase is characterized by $C_{{\rm SF}}(r)\sim r^{-K/2}$ with the Luttinger parameter $K<0.5$ for $\rho=1$ \cite{lut08, don01}. Both SS and DW possess a finite $\mathcal O_{{\rm DW}}=\lim_{r\to\infty}C_{{\rm DW}}(r)$, but SS is a gapless phase, and DW is a gapped one. Using the charge gap, which is the energy required for creating a particle-hole excitation \cite{tas21},
\begin{equation}
\Delta_L=E_L(N+1)+E_L(N-1)-2E_L(N),
\label{eg}
\end{equation}
we can distinguish between the gapless and gapped phases, where $E_L(N)$ is the ground-state energy of the ladder having $N$ bosons. In a finite system, $\Delta_L$ is non-zero irrespective of whether the phase is gapped or gapless. But the quantity $L\Delta_L$ for different $L$ coalesces for a gapless phase \cite{lut08}, and using this property, gapless and gapped phases are distinguished. As we discuss below, in our ladder setup, we find two kinds of gapless phases with a finite density wave order, one with $K<0.5$ and the other with $K>0.5$, term as supersolid-one (SS1)  and supersolid-two (SS2) phases, respectively. Finally, we use the correlation function,
\begin{equation}
\mathcal R= \frac{1}{L}\sum_i\left(\hat b_{i, 2}^{\dagger}\hat b_{i, 1}+\hat b_{i, 1}^{\dagger}\hat b_{i, 2}\right),
\end{equation}
to quantify the coherence between the rungs.

\section{Ground state phase diagrams}
\label{gsp}
In the following, we analyze the ground state phases for a fixed $U/J$, i.e., in the $t_\perp-V$ plane, for a small ($U/J=0.8$), moderate ($U/J=2$) and sufficiently large ($U/J=6$) onsite interactions. The three cases capture all quantum phases in a unit-filled EBHM ladder.

\subsection{Weak onsite interactions ($U/J=0.8$)}
\label{u08}
%


\begin{figure}
    \centering
    \includegraphics[width=\columnwidth]{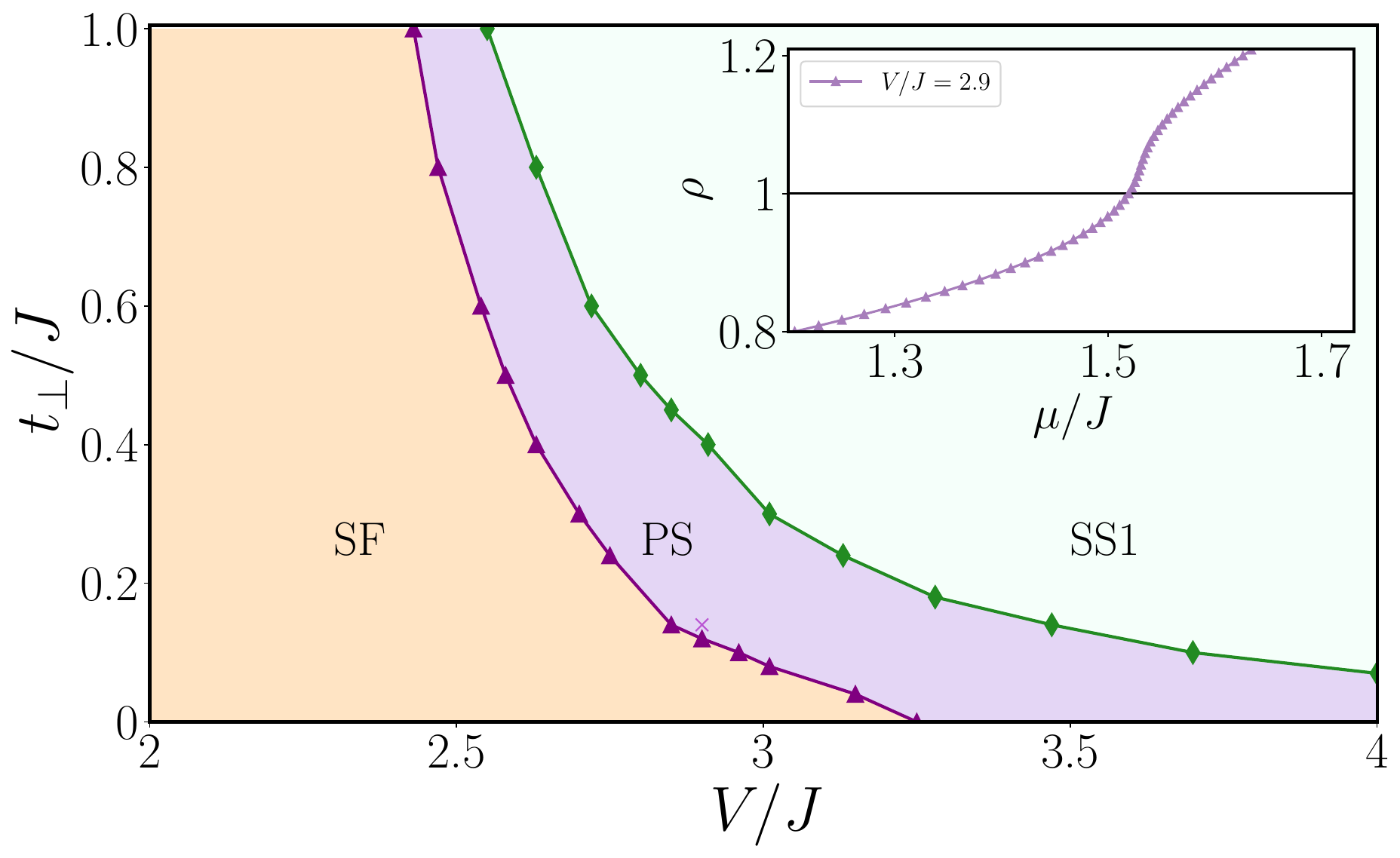}
    \caption{(Color online) The ground state phase diagram for $U/J=0.8$. The region of PS is identified by a discontinuous jump in the filling $\rho$ as a function of the chemical potential $\mu$ as shown in the inset for $t_\perp/J=0.14$. For calculating the phase boundaries, we used a maximum of six bosons per site ($n_b=6$), a bond dimension of 550, and $L=60$.} 
\label{fig:2}
\end{figure}
%

{\em Single chain} ($t_\perp=0$). The ground state is an SF for small $V/J$ [see Fig.~\ref{fig:2}]. As $V/J$ increases, the transition to phase separation occurs at $V/J\sim 3.25$, which is obtained by analyzing the dependence of filling $\rho$ on the chemical potential $\mu$ as detailed in Refs.~\cite{bat14, kot21,kra22}. In particular, the PS state is characterized by a discontinuous jump of $\rho$ as a function of $\mu$ across $\rho=1$ [see the inset of Fig.~\ref{fig:2}]. As $V/J$ increases to $\sim 5.66$, a transition to DW occurs, which is not shown in Fig.~\ref{fig:2}. The DW has an occupation pattern of 2 0 ... 0 2 in each chain i.e., $\langle \hat n_i\rangle=2$ in alternate sites.

\begin{figure}
    \centering
    \includegraphics[width=\columnwidth]{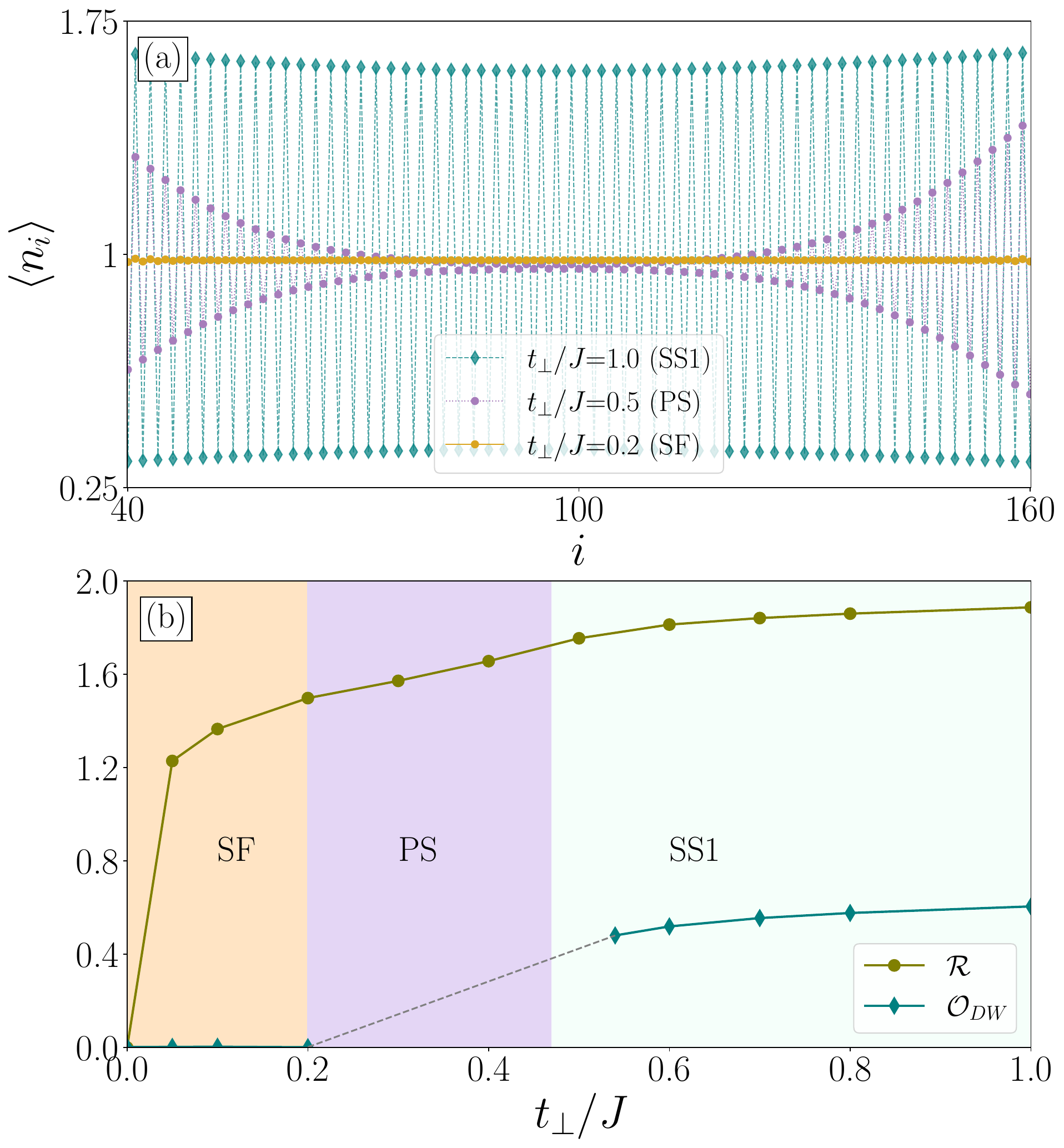}
    \caption{(Color online) (a) Occupation pattern $\langle \hat n_i \rangle$ of  SF ($t_\perp/J=0.2$), PS ($t_\perp/J=0.5$) and SS1 ($t_\perp/J=1$) phases for $V/J=2.6$ obtained using $L=151$. (b) DW order parameter $\mathcal O_{{\rm DW}}$ per site and the rung correlation $\mathcal R$ as a function of $t_\perp/J$ for $V/J=2.8$ and $U/J=0.8$, obtained after the finite size scaling. In the PS region, $\mathcal O_{{\rm DW}}$ is not well defined and hence, not shown. } 
\label{fig:2a}
\end{figure}

{\em Ladder} ($t_\perp\neq 0$). Introducing the rung hopping modifies the nature of phase transitions. PS is very robust against interchain hopping. Strikingly, supersolid SS1 emerges at intermediate values of $V/J$, which contrasts highly with the unit-filled single chain EBHM, where the supersolid does not exist. Thus, for a finite $t_\perp/J$, and increasing $V/J$, the ground state transitions from superfluid to phase-separation, supersolid, and finally to a gapped density wave phase. As the rung hopping amplitude $t_\perp/J$ increases, the SF-PS and PS-SS1 transition boundaries shift to smaller values of $V/J$. It implies that a finite $t_\perp/J$ can help induce density wave character in ground states for moderate values of $V/J$. For instance, while keeping $V/J=2.8$, increasing $t_\perp/J$ leads to a transition from SF to PS and then to SS1. The corresponding occupation patterns are shown in Fig.~\ref{fig:2a}(a), and the rung correlation $\mathcal R$ and DW order parameter $\mathcal O_{{\rm DW}}$ per site are shown in Fig.~\ref{fig:2a}(b). As expected, $\mathcal R$ increases with $t_\perp/J$ and is saturated at large $t_\perp/J$, and a finite DW order parameter is seen above a particular $t_\perp/J$. In the region of PS, the DW order parameter is not well defined and, hence, not shown in Fig.~\ref{fig:2a}(b). In the following sections, we discuss the SS1-DW transition in more detail when considering large values of $U/J$.

%
\subsection{Moderate on-site interactions ($U/J=2$)}
\label{u2}
%
\begin{figure}
    \centering
    \includegraphics[width=0.9\columnwidth]{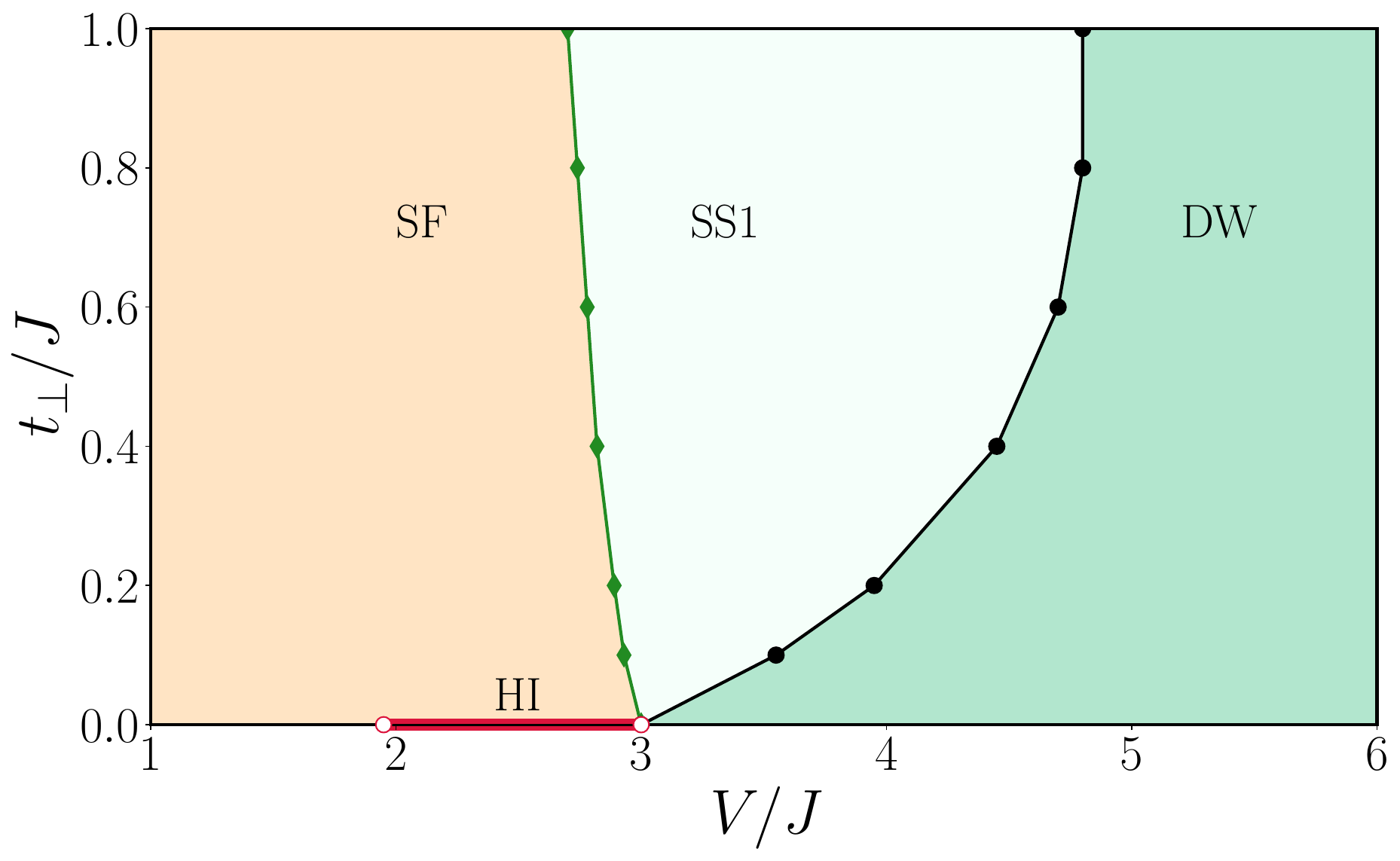}
    \caption{(Color online) Phase diagram for $U/J=2$. The HI emerges in a single ($t_\perp=0$) between $V/J \sim 1.95$ and $V/J\sim 3$, as shown by the solid brown line with open circles at the ends.} 
    \label{fig:3}
\end{figure}

%

\begin{figure}
    \centering
    \includegraphics[width=0.9\columnwidth]{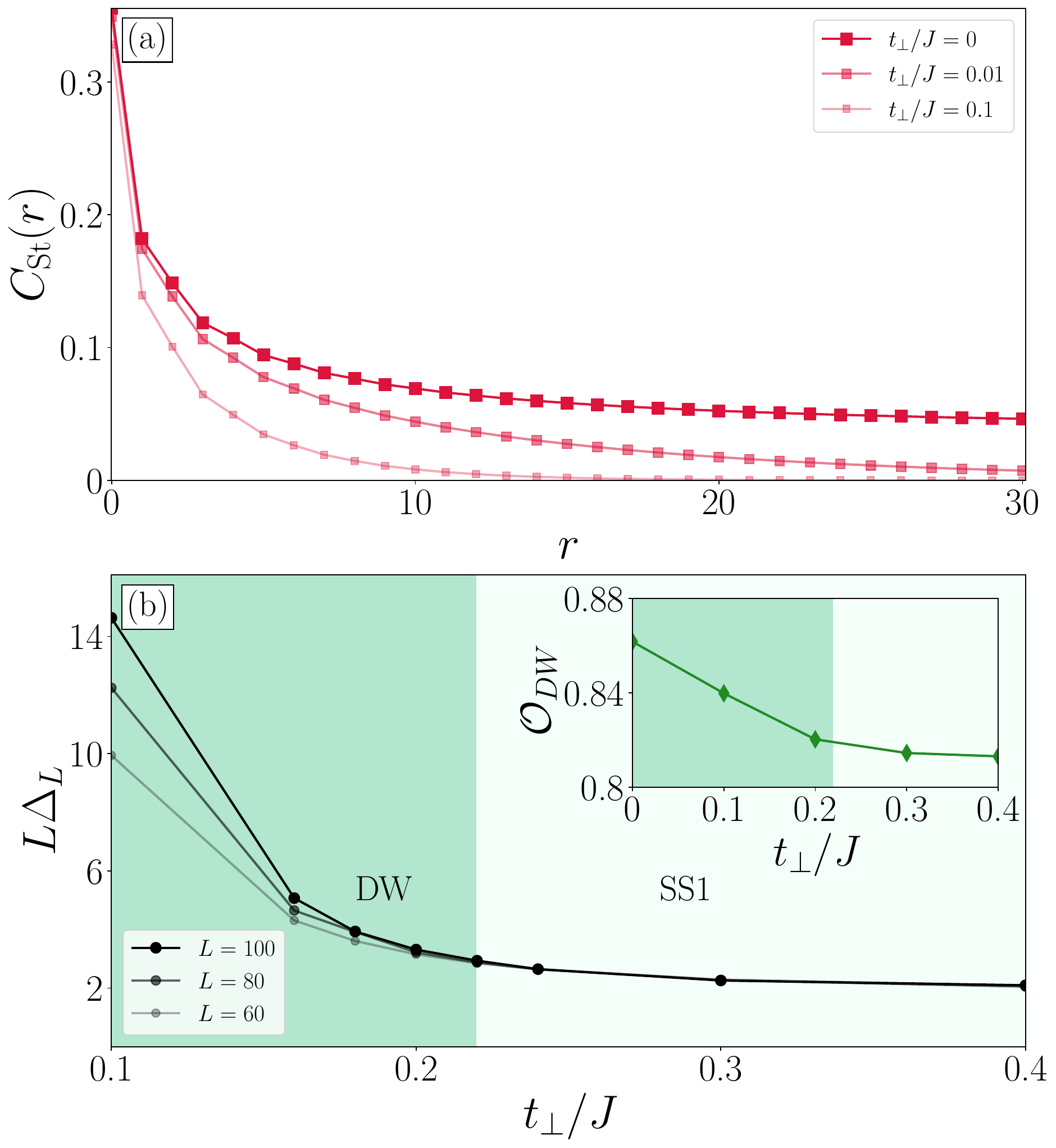}
    \caption{(Color online) (a) String correlation function $C_{{\rm St}}(r)$ for $U/J=2$ and $V/J=2.5$. For an isolated chain (filled circles), it converges to a non-zero value at large $r$, whereas in a ladder for a non-zero $t_\perp$ (filled triangles and squares), it decays to zero. (b) $L\Delta_L$ vs $t_\perp/J$ reveals the gapless-gapped transition for $U/J=2$ and $V/J=4.15$.  The DW order parameter in the inset is obtained by finite-size scaling.} 
    \label{fig:4}
\end{figure}

The ground state phase diagram for moderate onsite interactions ($U/J=2$) is shown in Fig.~\ref{fig:3} and is qualitatively different from that for weak onsite interactions. PS no longer appears in the phase diagram; instead, HI exhibiting a finite string order emerges at intermediate $V/J$ (from $\sim 1.95$ to 3) when the chains are disconnected ($t_\perp=0$). Hence, in 1D EBHM, the transition occurs from SF to HI and finally to DW as $V/J$ increases. HI is marked by a solid brown line along the horizontal axis in Fig.~\ref{fig:3}. An infinitesimally small $t_\perp$ destroys the string order and replaces HI with the gapless SF phase. As shown in Fig.~\ref{fig:4}(a), the string correlation function $C_{{\rm St}}(r)$ along a chain decays and vanishes for a finite $t_\perp/J$. Similar behavior of string order is reported in a spin-1 two-leg antiferromagnetic ladder \cite{tod01, anf07}. It is attributed to the nonlocal commutation between the string order parameter and the interchain hopping. 

As in the case of $U/J=0.8$, the SS1 phase emerges at finite $t_\perp$ and for sufficiently large $V/J$. Hence, a direct transition from SF to SS1 exists, and as $t_\perp$ increases, the SF-SS1 transition point shifts to a slightly smaller value of $V/J$. At larger $V/J$, we see the transition from SS1 to DW. The SS1-DW transition point shifts to a larger $V/J$ with an increase in $t_\perp/J$, eventually saturating at larger $t_\perp/J$. It indicates that it is possible to observe a transition from the gapped DW to a gapless SS1 by increasing $t_\perp/J$ while keeping $V/J$ constant. The gapped-gapless transition is demonstrated via $L\Delta_L$ vs $t_\perp/J$ plot for different system sizes, as shown in Fig.~\ref{fig:4}(b). As seen in the gapless phase, the curves for different $L$ coalesce \cite{lut08}. Interestingly, it also implies that, here, the DW character reduces with an increase in rung-hopping [see the inset of Fig.~\ref{fig:4}(b)], contrary to what we have seen across the SF-PS-SS1 transition discussed above in the case of weak onsite interactions (see Fig.~\ref{fig:2a}). 

Further, analyzing the superfluid correlation function, we see that across the DW-SS transition, the Luttinger parameter changes from $K>0.5$ (DW region) to $K<0.5$ (SS1 region). It implies that the DW-SS transition is a Kosterlitz-Thouless type \cite{kuh00, bat13, bat14}, identical to the MI-SF transition in a Bose-ladder with onsite interactions \cite{don01}. The latter arises from the competition between the Umklapp scattering favoring MI and the inter-chain hopping stabilizing the superfluid order between the two chains.

\subsection{Large onsite interactions ($U/J=6$)}
\label{loi}

\begin{figure}[hbt]
    \centering
    \includegraphics[width=\columnwidth]{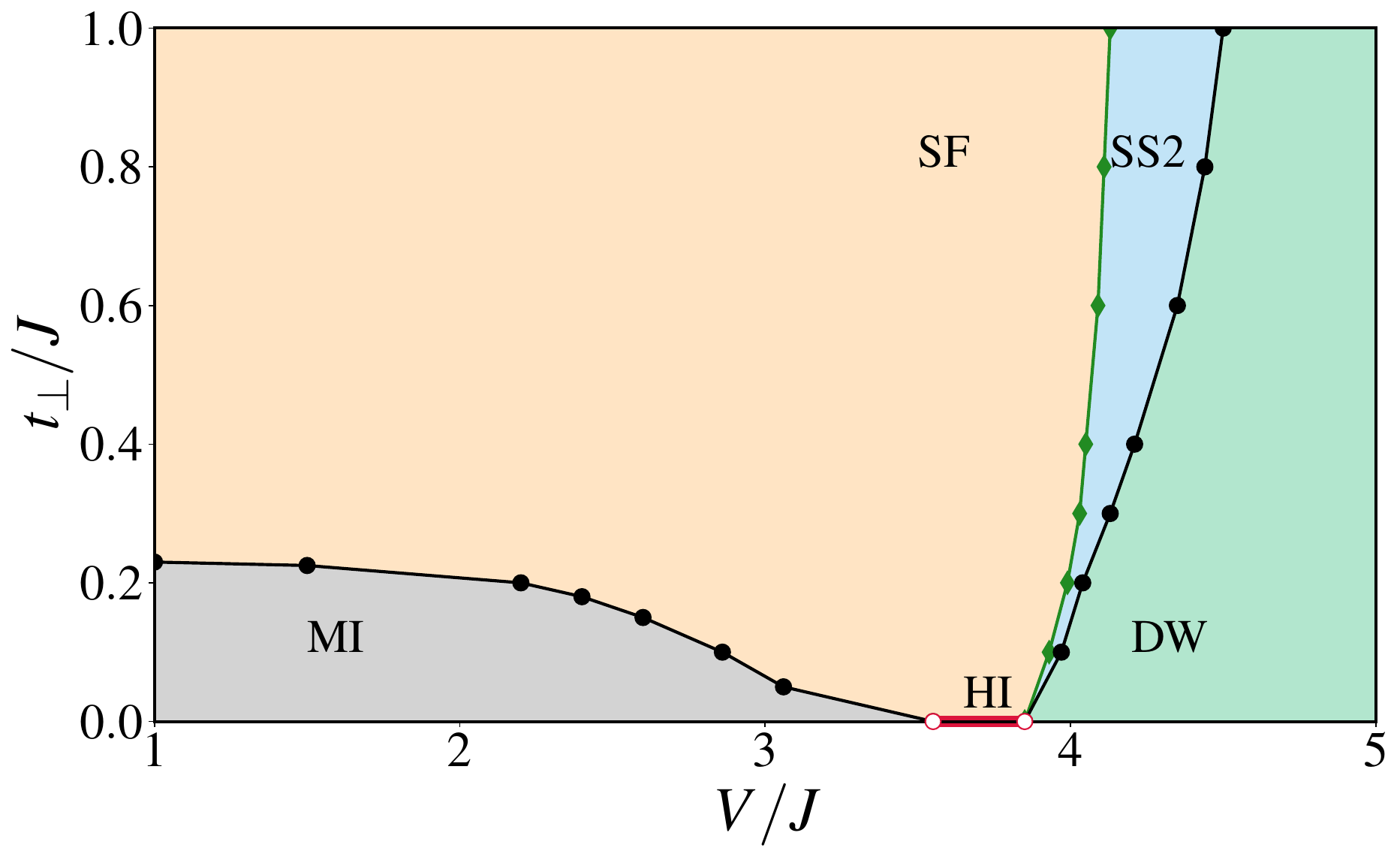}
    \caption{(Color online) Phase diagram for $U/J=6$. The HI exists for $3.55<V/J<3.85$ at $t_\perp=0$. } 
    \label{fig:5}
\end{figure}

Now, we consider a sufficiently large value of $U/J$ such that MI emerges in the ground state phase diagram [see Fig.~\ref{fig:5}]. In particular, MI appears at the small values of $t_\perp/J$ and $V/J$. At small $V/J$, increasing the rung hopping amplitude leads to MI-SF transition, as in a usual Bose-Hubbard ladder, which is a Kosterlitz-Thouless type transition \cite{don01}. The larger the value of $V/J$, the smaller $t_\perp/J$ is required for the MI-SF transition, resulting in an unusual scenario of getting a gapped (MI) to gapless (SF) by augmenting the density-density interactions. As before, the HI has not prevailed in the presence of rung hoppings and is replaced by SF. Unlike before, we found SS2 instead of SS1 here, and the SS region has significantly shrunk. A significant change occurs in the SF-SS2 boundary; instead of shifting to a lower value of $V/J$, it shifts to a larger value. It implies that at large values of onsite interactions, the rung-hoppings cannot induce a DW order starting from an SF phase. Hence, the DW order behaves differently with rung hoppings and strongly depends on density-density interaction. 

\subsection{Re-entrant transitions}
\label{rent}
\begin{figure}
    \centering
    \includegraphics[width=\columnwidth]{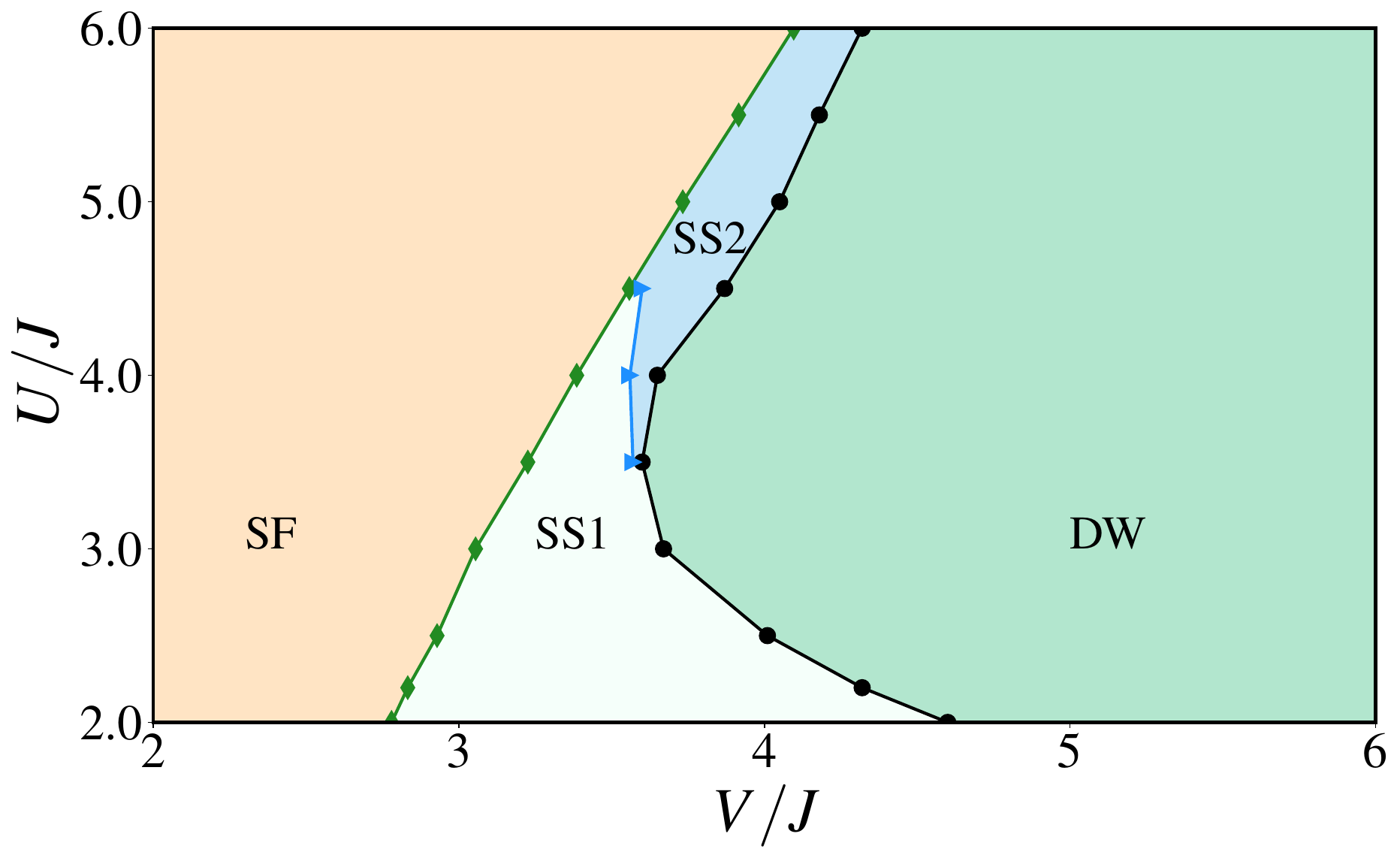}
    \caption{(Color online) Phase diagram for $t_{\perp}/J=0.6$. SS1 has a Luttinger parameter less than 0.5, and SS2 has a Luttinger parameter greater than 0.5, which is found to hardly change with the system size once $L$ is sufficiently large.} 
    \label{fig:6}
\end{figure}

Strikingly, the replacement of HI by a gapless SF for a finite $t_\perp$ leads to a re-entrant gapped transition as a function of $V/J$ in Fig.~\ref{fig:5}. For instance, taking $t_\perp/J=0.1$ and increasing $V/J$, the ground state changes from a gapped MI to a gapped DW via intermediate gapless (SF and SS2) phases. The re-entrant transition we discuss here differs from the previous reentrance behavior predicted in 1D BHM where only one kind of gapped (MI) phase is involved \cite{kuh98, pin12}.

\begin{figure}
    \centering
    \includegraphics[width=\columnwidth]{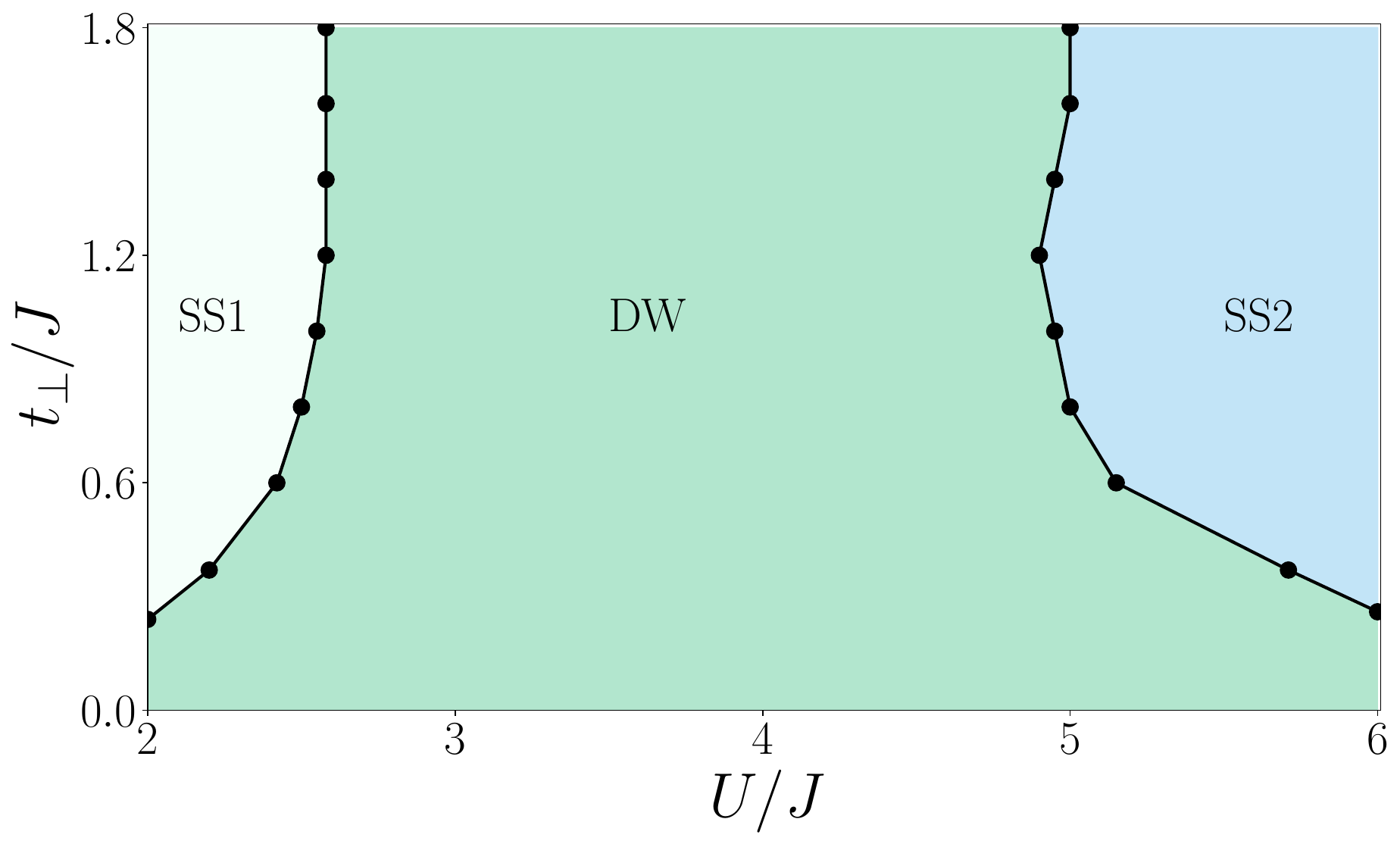}
    \caption{(Color online) Phase diagram in $t_\perp-U$ plane for $V/J=4.1$. SS1 is the supersolid phase with a Luttinger parameter less than 0.5 and SS2 is the same with a Luttinger parameter greater than 0.5.} 
    \label{fig:7}
\end{figure}

Interestingly, phase diagrams shown in Figs.~\ref{fig:6} and \ref{fig:7} also reveal reentrance behaviors, particularly a gapless one. In Fig.~\ref{fig:6}, we show the phase diagram in the $U-V$ plane for a fixed and sufficiently large $t_\perp$. We have an SF state for weaker inter-particle interactions ($U/J$ and $V/J$). Since the y-axis in Fig.~\ref{fig:6} is restricted to $U/J=6$ and $t_\perp/J$ is sufficiently large,  MI does not appear in the phase diagram. Thus, keeping $U/J$ constant and increasing the nearest neighbor interaction $V/J$, the ladder transitions from SF to SS and DW. The SS phase is SS1 for small $U/J$ and SS2 for large $U/J$. The non-monotonous behavior of the SS-DW boundary results in a reentrant gapless transition. In particular, at the intermediate values of $V/J$, there is a transition from SS1 (gapless) to DW (gapped) and then to SS2 (gapless). To reveal the robustness of this reentrant region, we obtain the phase diagram in $t_\perp-U$ plane for $V/J=4.1$ [see Fig.~\ref{fig:7}], which shows that the DW is sandwiched between SS1 and SS2 over a more extensive region. 

\section{Experimental Realizations}
\label{exp}

\begin{figure}
    \centering
    \includegraphics[width=\columnwidth]{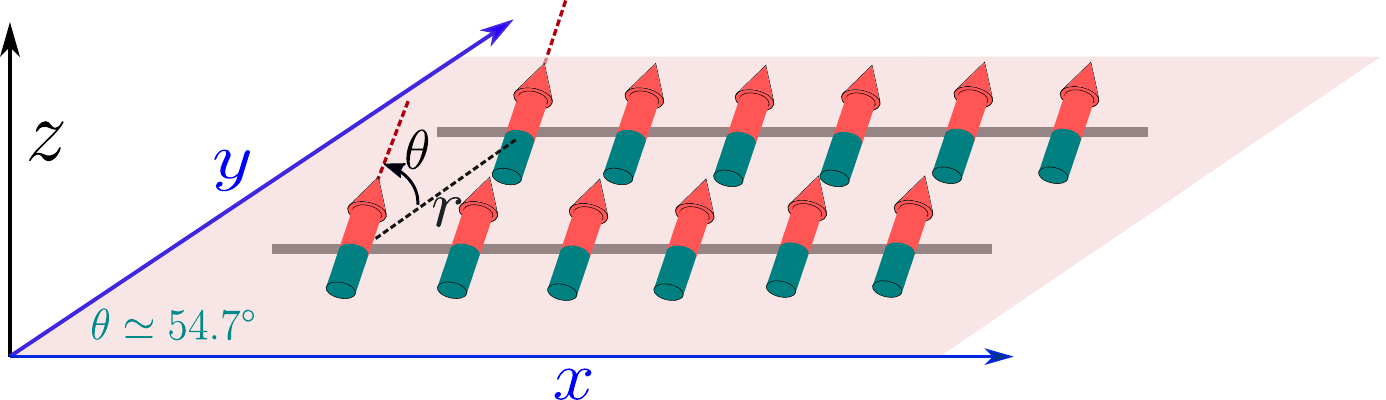}
    \caption{(Color online) Schematic diagram of the setup for implementing the Hamiltonian in Eq.~(\ref{ham}) using atomic dipoles or polar molecules.} 
    \label{fig:8}
\end{figure}

This section outlines two physical platforms where the Hamiltonian in Eq.~(\ref{ham}) can be experimentally probed. One is based on dipolar atoms \cite{bai16} or polar molecules \cite{big23, big24}, and the second uses Rydberg-admixed atoms loaded in optical lattices \cite{gla14, cho16, yij20}. The schematic setup for realizing the Hamiltonian in Eq.~(\ref{ham}) using dipoles is shown in Fig.~\ref{fig:8}. The ladder lies in the $xy$-plane with chains along the $x$-axis, and dipoles are oriented in $yz$ plane, forming an angle $\theta=54.7$ degrees (so-called magic angle) with $y$-axis \cite{sou19}. The angle is such that the dipoles do not interact along the $y$-axis, whereas they can hop in that direction. Along the $x$-axis, they repel each other. Then, truncating the dipole-dipole interactions beyond the nearest neighbor realizes the Hamiltonian in Eq.~(\ref{ham}). Note that the onsite interactions are provided by the usual $s$-wave scattering at low temperatures.

An alternative scheme is to use Rydberg-admixed atoms in ladder \cite{fro22}, and in particular, the Rydberg state is chosen to be a $nP_{3/2}$ state \cite{gla14}, where $n$ is the principal quantum number. Two Rydberg excited atoms separated by a distance of $r$, in $nP_{3/2}$ state interacts via an anisotropic interaction, 
\begin{equation}
V(r, \theta)\sim \frac{(ea_0)^4n^{11}}{r^6}\sin^4\theta=\frac{C_6(\theta)}{r^6},
\end{equation}
where $e$ is the electron charge, $a_0$ is the Bohr radius, and $\theta$ is the angle between the quantization axis (direction of the Zeeman field) and the radial vector between two atoms. For $\theta=0$, the two excited atoms do not interact, whereas, for $\theta=\pi/2$, they maximally interact and are repulsive. Again, we assume the ladder lies in the $xy$-plane, and the bosonic chains are along the $x$-axis. To emulate the Hamiltonian in Eq.~(\ref{ham}), we take the quantization axis to be along the $y$-axis; therefore, atoms along the rungs do not interact. Finally, admixing the electronic ground state of the bosonic atoms to a $nP_{3/2}$ state with a negative atom-field detuning $\Delta$ and Rabi frequency $\Omega$ such that $\Omega/|\Delta|\ll 1$, the atoms in the admixed state interacts via a two-atom potential \cite{hen10},
\begin{equation}
V_a(r, \theta)=\frac{\tilde C_6(\theta)}{r^6+R_c^6},
\end{equation}
where $\tilde C_6=(\Omega/2|\Delta|)^4C_6$ and $R_c=(C_6/2\hbar|\Delta|)^{1/6}$. The hopping along both chains and rungs can be independently controlled by the intensity of the laser fields forming the optical lattices in both schemes. We can also implement the same Hamiltonian using a single chain, but considering two energy bands and the atoms weakly admixed to a Rydberg $nS_{1/2}$ state \cite{cho16}.  
\section{Summary and outlook}
\label{cao}

In summary, we analyzed the effect of inter-chain hopping on the ground state phases of an extended Bose-Hubbard ladder for unit-filling fraction using finite-size DMRG. Contrary to the 1D chain, we find a supersolid phase. Our analysis reveals that the string correlations do not survive when there is a hopping between the chains. Interestingly, reentrant gapless and gapped transitions are found. Finally, we discussed two experimental platforms to probe our findings: dipolar atoms or polar molecules and Rydberg admixed atoms in optical lattices. One interesting aspect for future studies would be to analyze the quench dynamics across the various transitions, particularly by linearly quenching the hopping between the chains and looking for universal behaviors \cite{yij20, sab21}.

\section{Acknowledgements} 
We acknowledge funding from the EPSRC through Grant No. EP/W015641/1, the Going Global Partnerships Programme of the British Council (Contract No. IND/CONT/G/22-23/26), DST-SERB for the Swarnajayanti fellowship (File No. SB/SJF/2020-21/19) and MATRICS Grant No. MTR/2022/000454 from SERB, Government of India. We further  thank the National Supercomputing Mission for providing computing resources of ``PARAM Brahma'' at IISER Pune, which is implemented by C-DAC and supported by the Ministry of Electronics and Information Technology and Department of Science and Technology (DST), Government of India, and acknowledge National Mission on Interdisciplinary Cyber-Physical Systems of the Department of Science and Technology, Government of India, through the I-HUB Quantum Technology Foundation, Pune, India. This research is also funded in part by the Austrian Science Fund (FWF) [P 36850-N].
\bibliographystyle{apsrev4-1}
\bibliography{libladd.bib}
\end{document}